\documentclass[fleqn,twoside]{article}

\usepackage{
color}
\usepackage{espcrc2}

\usepackage{graphicx}
\usepackage[figuresright]{rotating}

\newcommand{\AmS}{{\protect\the\textfont2
  A\kern-.1667em\lower.5ex\hbox{M}\kern-.125emS}}

\title{Hard exclusive production of a vector meson}

\author{
D.\,Yu.\,Ivanov\address[IKS]{Sobolev Institute of Mathematics RAS,
 630090 Novosibirsk, Russia}
        \thanks{
D.~I. is supported by
grants DFG 436, RFBR 03-02-17734.
},
        G.\, Krasnikov\address{Department of Theoretical Physics,
St.Petersburg State University},
       and L.\, Szymanowski\address{
So{\l}tan Institute for Nuclear Studies, Hoza 69, 00-681 Warsaw, Poland}}
       
\begin{document}

\begin{abstract}
The processes of a light neutral vector meson, 
$V=\rho^0, \omega, \phi$, electroproduction and a heavy quarkonium, 
$V=J/\Psi, \Upsilon $, photoproduction are studied in the framework of QCD factorization. 
We derive a complete set of hard-scattering
amplitudes which describe these processes at next-to-leading order (NLO). 
\vspace{1pc}
\end{abstract}

\maketitle

\section{INTRODUCTION}
The strong interest in the processes of
elastic neutral vector meson electroproduction on a nucleon,
\begin{equation}
\gamma^*(q)\, N(p) \to V(q^\prime)\, N(p^\prime)
\, ,
\label{process}
\end{equation}
where $V=\rho^0,\,  \omega, \, \phi $, 
or heavy quarkonium $V=J/\Psi, \Upsilon$, is related with 
the possibility to constrain the gluon density in
a nucleon \cite{Ryskin:1992ui,BFGMS94}.

The large negative virtuality of the photon,
$q^2=-Q^2$, or (in a case of quarkonium
photoproduction) the heavy quark mass, $m$, provides a hard
scale for the process which justifies the application of QCD factorization
methods that allow to separate the contributions to the amplitude
coming from different scales.
The amplitude of light vector meson electroproduction 
is given by a convolution of the
nonperturbative
meson distribution amplitude (DA) and the generalized parton densities
(GPDs)
with the perturbatively calculable hard-scattering amplitudes \cite{CFS96}.

The direct application of factorization theorem \cite{CFS96} 
to the production of a heavy meson is restricted
to
the region of very large virtualities, $Q^2\gg m^2$, where the mass of the
heavy quark may be completely neglected. In contrast,
in photoproduction
or electroproduction at moderate virtualities the heavy quark
mass provides a hard scale
and the nonrelativistic nature of heavy meson is important.
In this case, according to nonrelativistic QCD (NRQCD) which
provides a systematic nonrelativistic expansion, a factorization formalism
must be constructed in terms of matrix elements of NRQCD operators. They are
characterized by their different scaling
behavior with respect to $v$, the typical
velocity of the heavy quark. In the leading approximation only the
matrix element $\langle O_1
\rangle_V$ contributes, which describes in NRQCD the leptonic meson decay
rate \cite{Bodwin:1994jh}
\begin{equation}
\Gamma[V\to l^+l^-]=\frac{2e_q^2\pi\alpha^2}{3}
\frac{\langle O_1\rangle_V }{m^2}
\left( 1-\frac{8\alpha_S}{3\pi}\right)^2 \, .
\label{decay}
\end{equation}
Here $\alpha$ is the fine-structure constant and $m$ and $e$ are
the pole mass and the electric charge of the heavy quark ($e_c=2/3$,
$e_b=-1/3$) and $\alpha_S$ is the strong coupling
constant.

In this contribution we present our recent results for
the hard scattering amplitudes at
next-to-leading order both for heavy quarkonium photoproduction
\cite{heavy} and for light vector meson electroproduction \cite{rho}
processes. The account of the radiative correction to the hard scattering
amplitudes allows to reduce the theoretical uncertainty of the
factorization approach, the
dependence on the factorization, $\mu_F$, and renormalization, $\mu_R$,
scales, which is especially
important at high energies, since in this case (i.e. in the
small $x$ region) the
dependence of the parton distributions on the factorization
scale is very strong.

\section{QUARKONIUM PRODUCTION}

In the leading order of the
relativistic expansion  the meson mass can be taken
as  twice the
heavy quark pole mass, $(q^\prime)^2=M^2$ and $M=2m$.
$p^2=p^{\prime 2}=m_N^2$, where $m_N$ is the proton mass.
The photon
polarization is described by the vector
$e_\gamma$, $(e_\gamma q)=0$.
The invariant c.m. energy is $s_{\gamma p}=(q+p)^2=W^2$. We define
\begin{eqnarray}
&&
\Delta=p^\prime -p \, , \ \ P=\frac{p+p^\prime}{2} \, , \ \ t=\Delta^2 \, ,
\nonumber \\
&&
(q-\Delta )^2=(q^\prime)^2=M^2 \, , \ \ \zeta =\frac{M^2}{W^2} \, ,
\label{not1}
\end{eqnarray}

At $|t|\ll M^2$ the factorization formula  reads
\begin{eqnarray}
&&
 {\cal M}
=\frac{4\pi \sqrt{4\pi\alpha}\, e_q (e^*_V e_\gamma )}
{N_c \, \xi}\left(\frac{\langle O_1 \rangle_V}{m^3}\right)^{1/2}
 \int\limits^1_{-1} dx
\times \nonumber
\\
&&
\left[\, T_g( x,\xi)\, F^g(x,\xi,t)+
T_q (x,\xi) F^{q,S} (x,\xi,t) \,
\right] \, ,
\nonumber \\
&&
F^{q,S} (x,\xi,t)=\sum_{q=u,d,s}  F^q (x,\xi,t) \, .
\label{fact1}
\end{eqnarray}
as the sum of the gluon, $F^g(x,\xi,t)$, and the quark, 
$F^q (x,\xi,t)$, GPDs contributions. GPDs are defined as a functions which
parametrized the matrix elements of the renormalized 
light-cone quark and gluon
operators. The polarization vector of quarkonium is 
$e_V$, variable $\xi=\zeta/(2-\zeta)$ parametrizes the non vanishing 
longitudinal momentum transfer in the process.  

The hard scattering amplitude $T_g( x,\xi)$ (or $T_q (x,\xi)$) represents 
essentially the on-shell parton amplitude for the scattering of a 
pair of gluons (quarks) which are collinear to the proton momentum and have  
the fractions $(x+\xi)/(1+\xi)$ and $(x-\xi)/(1+\xi)$. 
Calculated in the dimensional regularization method these one-loop 
amplitudes contain
poles, the infrared collinear and the ultraviolet singularities. The full 
renormalization procedure includes mass counterterm diagrams, the
renormalization of the heavy quark field and the renormalization of the
strong coupling. The factorization of collinear singularities is achieved
by the replacement of the bare GPDs by the renormalized ones. This
procedure leads to the finite results for $T_g( x,\xi)$ and $T_q (x,\xi)$
at NLO  
\begin{eqnarray}
&&
T_q(x,\xi)=\frac{\alpha_S^2(\mu_R) C_F}{2\pi}
f_q\left(\frac{x-\xi+i\varepsilon}
{2\xi}\right) \, ,
\\
&&
T_g(x,\xi) =
 \frac{\xi}{(x-\xi+i\varepsilon)(x+\xi-i\varepsilon)}
\times  \\
&& 
 \left[
  \alpha_S(\mu_R) + \frac{\alpha_S^2(\mu_R)}{4\pi}
f_g\left(\frac{x-\xi+i\varepsilon}
{2\xi}\right)
 \right] \, ,
\nonumber
\label{TTg}
\end{eqnarray}
here $C_F=4/3$. The functions $f_q$ and $f_g$ (see \cite{heavy}) contain 
terms $\sim \ln(m^2/\mu_F^2)$.   

The dependence of NLO hard scattering amplitudes on $\mu_F$ 
compensates partially the effect of the evolution of GPDs with factorization
scale (the dependence of $T_g, T_q$ and GPDs on $\mu_F$ in (\ref{fact1}) 
is not
shown for shortness). That leads to the substantial reduction of the scale
ambiguity of the theoretical predictions in NLO in comparison with leading
order (LO), see Fig.~1

\begin{figure}[h]
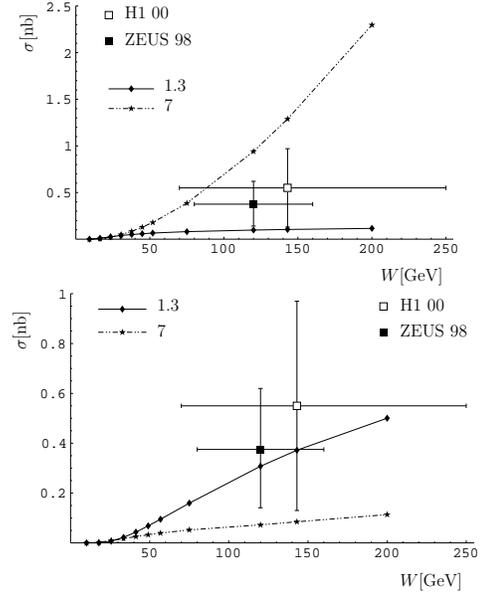

\label{5}
\begin{center}
\scalebox{0.55}{
\input{fig13.pstex_t}
}\,\,\,\,
\scalebox{0.55}{
\input{fig14.pstex_t}
}
\caption[*]{
The cross section of the $\Upsilon$ photoproduction;
the predictions at LO (upper figure) and NLO (lower figure)
for the scales
$\mu_F=\mu_R=[1.3,7]\mbox{ GeV}$.
The data are from ZEUS \cite{Breitweg:1998ki} and H1 \cite{Adloff:2000vm}.
For the $t-$ dependence we assumed exponential with the slope parameter
$b=4.4 \mbox{ GeV}^{-2}$.}
\vspace*{-0.8cm}
\end{center}
\end{figure}

\section{LIGHT MESON PRODUCTION}

We refer the reader to  \cite{rho,Ivanov:2004bw} for 
the complete set of analytical results and for some numerical predictions 
for the light vector meson electroproduction at NLO. 
The new features in comparison to quarkonium production are the contribution
of the quark GPD at LO and the appearance of 
a light meson DA in the factorization formulae (instead of NRQCD matrix
element).

Similarly to the quarkonium production case we observe that in HERA
kinematic range the NLO corrections are large and 
mostly of the opposite signs than the corresponding Born terms, consequently
the final values of the amplitude are the result of a strong 
cancelation between the LO and NLO parts. 
Leading contribution to the NLO correction comes from the
integration region $\xi\ll |x|\ll 1$, simplifying
the gluon hard-scattering amplitude in this limit we obtain 
the estimate
\begin{equation}
 {\cal M}_{\gamma^*_L N\to V_L N}
\approx  \label{appr}
\end{equation}
\vspace*{-0.3cm}
\begin{eqnarray}
 \frac{-2\, i\, \pi^2 \sqrt{4\pi\alpha}\, \alpha_S f_V Q_V}
{N_c \, Q \, \xi}\int\limits^1_{0}\frac{ dz \, \phi_V(z)}{z(1- z)}\Biggl[
F^g(\xi,\xi,t)  &&
\nonumber \\
+\frac{\alpha_S N_c}{\pi}
\ln\left(\frac{Q^2z(1-z)}{\mu_F^2}\right)
\int\limits^1_{\xi} \frac{dx}{x}  F^g(x,\xi,t)
%+ T_{(+)} ( z, x)  F^{(+)} (x,\xi,t)
\Biggr] \, , \nonumber &&
\end{eqnarray}
here  $f_V$ is a meson coupling constant known from $V\to e^+e^-$ decay,
$N_c=3$, factor $Q_V$ depends on the meson flavor content \cite{rho},
$\xi=x_{Bj}/(2-x_{Bj})$.
Given the behavior of the gluon GPD at small $x$, $F^g(x,\xi,t)\sim const$,
we see that NLO correction is parametrically
large, $\sim \ln(1/\xi)$, and negative unless one chooses the
value of the factorization scale sufficiently lower than the kinematic
scale. For the asymptotic form of meson DA, $\phi^{as}_V(z)=6z(1- z)$,
the last term in (\ref{appr}) changes the sign at $\mu_F=\frac{Q}{e}$,
for the DA with a more broad shape this happens at even lower values of
$\mu_F$.

\begin{figure}[h]
\label{2}
\vspace*{0.0cm}
\begin{center}
%\centerline{\epsfxsize=2.4in\epsfbox{2i5gev.eps}}
\includegraphics[scale=0.7]{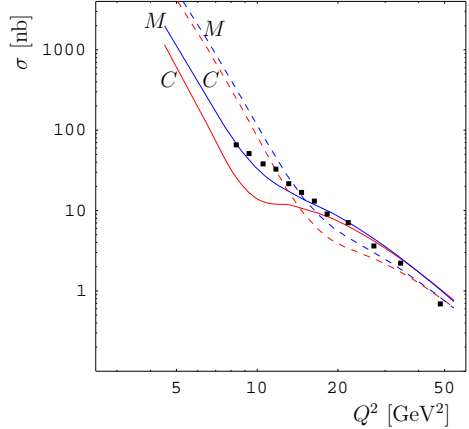}
\vspace*{-1cm}
\caption[*]{ $\sigma(Q^2,W=95\mbox{GeV})$ as a function
of $Q^2$ for two NLO GPDs:
MRST2001 (M) and CTEQ6M (C).
The factorization scale
$\mu_F=Q$. For the solid lines
 $\mu_R=\mu_F$, for the dashed lines
$\mu_R=Q/\sqrt{e}$.
The data points are taken from  \cite{Kreisel}.
}
\vspace*{-1cm}
\end{center}
\end{figure}

Our leading twist results, see Fig.~2, obtained with NLO
hard-scattering amplitudes and NLO GPDs \cite{Freund:2002qf} (
which were adjusted to describe
HERA deeply virtual
Compton scattering data)
are in qualitative agreement with the measured at HERA $\rho$ meson
electroproduction cross section.      
Without account of NLO terms the
predictions would be substantially above the data.


\begin{thebibliography}{9}

\bibitem{Ryskin:1992ui}
M.\,G. Ryskin,
%Diffractive J / psi electroproduction in LLA QCD,''
Z.\ Phys.\ C {\bf 57}, 89 (1993).

\bibitem{BFGMS94}
S.\,J. Brodsky at al.,
%Diffractive leptoproduction of vector mesons in QCD,''
Phys.\ Rev.\ D {\bf 50}, 3134 (1994).

\bibitem{CFS96}
J.\,C. Collins, L. Frankfurt and M. Strikman,
%Factorization for hard exclusive electroproduction of mesons in
Phys.\ Rev.\ D {\bf 56}, 2982 (1997).

%\cite{Bodwin:1994jh}
\bibitem{Bodwin:1994jh}
G.T.~Bodwin, E.~Braaten and G.P.~Lepage,
%Rigorous QCD analysis of inclusive annihilation and production of heavy
%quarkonium,''
Phys.\ Rev.\ D {\bf 51}, 1125 (1995)
[Erratum-ibid.\ D {\bf 55}, 5853 (1997)].
%%CITATION = HEP-PH 9407339;%%

 \bibitem{heavy}
%\cite{Ivanov:2004vd}
%\bibitem{Ivanov:2004vd}
D.~Yu.~Ivanov, A.~Sch{\"a}fer, L.~Szymanowski and G.~Krasnikov,
%`Exclusive photoproduction of a heavy vector meson in QCD,''
Eur.\ Phys.\ J.\ C {\bf 34} (2004) 297.
%%CITATION = HEP-PH 0401131;%%

\bibitem{rho}
%\cite{Ivanov:2004vd}
%\bibitem{Ivanov:2004vd}
D.~Yu.~Ivanov, L.~Szymanowski and G.~Krasnikov,
%ector meson  electroproduction at next-to-leading order,''
 JETP Lett. {\bf 80} (2004) 255.


%\cite{Breitweg:1998ki}
\bibitem{Breitweg:1998ki}
J.~Breitweg {\it et al.}  [ZEUS Collaboration],
%Measurement of elastic Upsilon photoproduction at HERA,''
Phys.\ Lett.\ B {\bf 437}, 432 (1998).
%%CITATION = HEP-EX 9807020;%%

%\cite{Adloff:2000vm}
\bibitem{Adloff:2000vm}
C.~Adloff {\it et al.}  [H1 Collaboration],
%Elastic photoproduction of J/psi and Upsilon mesons at HERA,''
Phys.\ Lett.\ B {\bf 483}, 23 (2000).
%%CITATION = HEP-EX 0003020;%%

%\cite{Ivanov:2004bw}
\bibitem{Ivanov:2004bw}
D. Yu.~Ivanov, G.~Krasnikov and L.~Szymanowski,
%`Exclusive production of vector mesons in QCD,''
arXiv:hep-ph/0409110.
%%CITATION = HEP-PH 0409110;%%

\bibitem{Freund:2002qf}
A.~Freund, M.~McDermott and M.~Strikman,
Phys.\ Rev.\ D {\bf 67}, 036001 (2003).

\bibitem{Kreisel}
A.~Kreisel {\it et al.} [H1 Collaboration], arXiv: hep-ex/0208013.

\end{thebibliography}
\end{document}